\documentclass[pdflatex,sn-nature]{sn-jnl}

\usepackage[utf8]{inputenc}
\usepackage{newunicodechar}
\newunicodechar{∼}{\textasciitilde}
\usepackage{graphicx}%
\usepackage{multirow}%
\usepackage{amsmath,amssymb,amsfonts}%
\usepackage{amsthm}%
\usepackage{mathrsfs}%
\usepackage[title]{appendix}%
\usepackage{xcolor}%
\usepackage{textcomp}%
\usepackage{manyfoot}%
\usepackage{booktabs}%
\usepackage{algorithm}%
\usepackage{algorithmicx}%
\usepackage{algpseudocode}%
\usepackage{listings}%
\usepackage[section]{placeins}

\theoremstyle{thmstyleone}%
\theoremstyle{thmstyletwo}%

\theoremstyle{thmstylethree}%

\begin{document}

\title[Article Title]{Synaptic Multi-Weight Wafer-Scale Domain Wall Magnetic Tunnel Junctions Driven by Spin Orbit Torque}

\author[1, 2]{\fnm{Nicholas} \sur{Zogbi}}\email{nzogbi@utexas.edu}
\author[3,4]{\fnm{Domenico} \sur{Giuliano}}\email{domenico.giuliano@imec.be}
\author[3]{\fnm{Vaishnavi} \sur{Kateel}}\email{vaishnavi.kateel@imec.be}
\author[3]{\fnm{Van Dai} \sur{Nguyen}}\email{van.dai.nguyen@imec.be}
\author[3]{\fnm{Jyotirmoy} \sur{Chatterjee}}\email{jyotirmoy.chatterjee@imec.be}
\author[3]{\fnm{Alvaro} \sur{Palomino}}\email{alvaro.palominolopez@imec.be}
\author[3]{\fnm{Robert} \sur{Carpenter}}\email{robert.carpenter@imec.be}
\author[3]{\fnm{Siddharth} \sur{Rao}}\email{siddharth.rao@imec.be}
\author*[1,2]{\fnm{Jean Anne C.} \sur{Incorvia}}\email{incorvia@austin.utexas.edu}

\affil[1]{\orgdiv{Chandra Family Department of Electrical and Computer Engineering}, \orgname{University of Texas at Austin}}
\affil[2]{\orgdiv{Microelectronics Research Center}, \orgname{University of Texas at Austin}}
\affil[3]{IMEC, Kapeldreef 75, 3001 Leuven, Belgium}
\affil[4]{Department of Physics and Astronomy, Quantum Solid-State Physics (QSP) Division, Katholieke Universiteit Leuven, Celestijnenlaan 200D Box 2414, 3001 Leuven, Belgium}


\abstract{Spintronic synaptic devices based on magnetic random access memory (MRAM) are a rapidly evolving class within the analog neuromorphic architecture. However, obtaining the multi-weight or analog behavior desired for improved neural network inference accuracy in scaled MRAM remains a challenge. In this study, we demonstrate and tune up to eight distinct resistance levels in 300 mm wafer-scale magnetic tunnel junction artificial synapses driven by spin orbit torque. By varying the applied in-plane field, multi-weight resistance states emerge through domain wall nucleation and pinning. These multi-weight states appear stochastically and can be addressed with an iterative programming algorithm to program a specific resistance. Electromagnetic simulations indicate the multiple states are due to domain wall nucleation mediated by in-plane field and applied voltage. Implementation of the measured multi-weight devices in a neural network yields a 96.1\% inference accuracy using the MNIST dataset if all devices provide the same number of multi-weight states. When experimental device-to-device variation is included, the accuracy is 48.6\%. Overall, this work demonstrates a route towards multi-weight artificial synapse MRAM, establishing a path towards scalable neuromorphic systems based on spintronics.}


\maketitle


\section{Introduction}\label{intro}

Neural networks (NNs) have become a foundation of modern computing, powering various applications such as natural language processing, computer vision, and autonomous systems. However, their growing computational demands pose significant challenges of high energy consumption and worsens with increasing network size \cite{Strubell_2020, Davenport_2024}. Considering these challenges, deploying NNs at both the cloud and edge necessitates a shift towards more energy-efficient architectures. 

Digital implementations of NNs separate computation and memory, resulting in higher energy and area costs \cite{Bai_2017}. Analog neuromorphic architectures aim to embed computation and memory within the same device, increasing energy efficiency, improving parallelism, and enabling dense architectures such as crossbar arrays \cite{Amirsoleimani_2020}. This paves the way for designing hardware that directly mimics the behavior of neurons and synapses to efficiently perform NN operations such as matrix vector multiplication \cite{Indiveri_2011, Pazos_2025}. Current implementations of synaptic devices include resistive RAM (RRAM) \cite{Ielmini_2025, Woo_2016, Sung_2018, Moon_2016, Oh_2017, Esmanhotto_2022}, phase-change memory (PCM) \cite{Moraitis_2018, Suri_2011, Nandakumar_2020, Sarwat_2022, Ortner_2025, Joshi_2020},  ferroelectric RAM (FRAM) \cite{Cao_2025, Tang_2025, Cho_2025}, and electrochemical RAM \cite{Faruque_2025, Kireev_2022}. Each class of synaptic devices offers trade-offs between endurance, retention, switching speed, and analog programmability, shaping their suitability for different neuromorphic applications. Generally, a high number of analog weights is desirable: reducing the number of bits for synaptic weights limits network expressivity by quantizing and distorting pre-trained weights, which in turn degrades inference accuracy. Studies on limited numerical precision have shown that networks require a high bit-depth to preserve accuracy, reflecting the intrinsic sensitivity of NN performance to weight quantization \cite{Gupta_2015, Courbariaux_2015}. This highlights the importance of designing synaptic devices that support multi-weight (MW), and ideally analog, resistance states.

Spintronic devices are a rapidly evolving class of nonvolatile memories used as synaptic devices, including magnetic random-access memory (MRAM) based on a magnetic tunnel junction (MTJ). Leveraging magnetic switching mechanisms such as spin transfer torque (STT), spin orbit torque (SOT), and/or voltage-controlled magnetic anisotropy (VCMA), spintronic-based synapses offer non-volatility and high cycling endurance (e.g. 10$^{14}$) \cite{Worledge_2024}. Because of this high endurance, if there was a way to have true analog MRAM combined with a larger read window, MRAM could be a major contender for analog crossbar arrays. A major challenge of spintronic synapses is their natural bistability, preventing the ideal MW or analog behavior desired for high accuracy in crossbar array architectures. Numerous efforts have been made to design and prototype MW spintronic synapses, e.g. via multi-domain nucleation \cite{Jeong_2024, Liu_2022_compensatedferrimagnet}, domain walls (DWs) \cite{Liu_2021, Liu_2024, Koh_2025, Siddiqui_2019, Alam_2024}, skyrmions \cite{Huang_2017, song2020skyrmion, da2025neuromorphic}, and magneto-ionics \cite{bernard2025dynamic, liu2022bayesian, das2025magneto, monalisha2023multilevel, monalisha2024magnetoionics}. But, so far, the general assumption is that industrially-relevant wafer-scale MRAM is limited to binary states. This has motivated several approaches, including the use of binary MTJs as artificial synapses \cite{Jung_2022, Goodwill_2022_BNN, Liu_2025, doevenspeck2020sot}, encoding synaptic weights in alternative variables such as the RF frequency response \cite{Ross_2023, menshawy2026remotely}, and representing weights through a common electrode shared across multiple MTJs \cite{cai2022selective, doevenspeck2021multi}. These solutions either add circuit or device complexity \cite{nahshan2021loss, sehgal2025energy}, and wafer-scale MW or analog MRAM is missing.

In this study, we design and measure rectangular MRAM as a path toward MW and analog behavior. We demonstrate MW resistance states in 300 mm wafer MTJ artificial synapses driven by SOT. While a 60 nm diameter circular MTJ stays binary, increasing the aspect ratio $\sim$5x between the length and width of the MTJ results in MW states. Statistical analysis across the wafer is performed, showing the MW behavior is tunable via applied in-plane (IP) magnetic field. This results in a single bit weight at higher fields and a maximum of three bits at lower fields. Although each weight state is not obtained every cycle, we show that through cycling the weight can be programmed, making MTJs suitable for back propagation training. Electromagnetic simulations are performed, indicating that domain wall nucleation occurs due to leakage current from the larger aspect ratio creating a potential difference between the SOT track and edges of the MTJ top electrode. Implementation of the measured device-to-device variations in a NN shows a 48.6\% inference accuracy on the given tasks, increasing to 96.1\% inference accuracy if all devices operate equally. This work establishes a pathway towards scalable analog spintronic devices for neuromorphic systems.


\section{Device Structure}\label{device_structure}

\begin{figure}[htbp]
    \centering
    \includegraphics[width=1\textwidth]{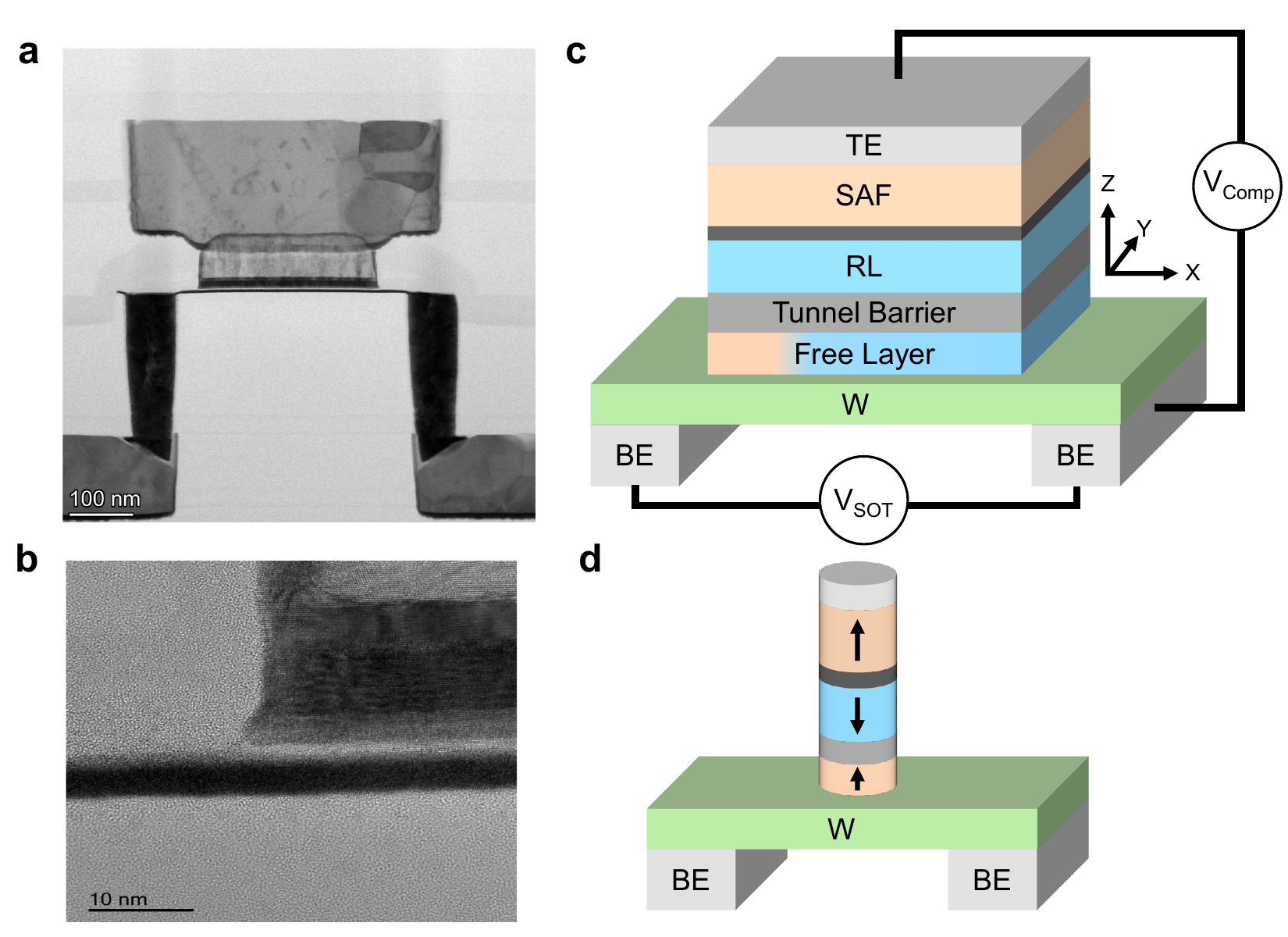}
    \caption{\textbf{MTJ artificial synapses integrated on a 300 mm wafer.} (\textbf{a}) Cross-sectional TEM image of the MTJ artificial synapse with a length of 350 nm. (\textbf{b}) Cross-sectional TEM of the MTJ artificial synapse's edge profile. (\textbf{c}) Schematic of SOT-MTJs with varying lengths of [350 nm, 400 nm, and 600 nm] and widths of [70 nm, 90 nm, and 120 nm]. The tungsten heavy metal is depicted as W; BE, RL, SAF, and TE indicate the bottom electrode, reference layer, synthetic anti-ferromagnet, and top electrode of the MTJ, respectively. Orange and blue regions represent opposite magnetization directions. (\textbf{d}) Schematic of 60 nm diameter SOT-MTJ control.}
    \label{fig1}
\end{figure}

Four different SOT-MTJ configurations with perpendicular magnetic anisotropy were fabricated on a $\beta$-W heavy metal track with a CoFeB free layer and MgO tunnel barrier, as described in Ref. \cite{Kateel_2023}, with details discussed in Section \ref{methods}. In total, 128 devices were tested on a 300 mm wafer, with 32 devices per configuration, enabling wafer-scale statistical analysis. A cross-sectional TEM of a 350 nm artificial synapse is shown in  Fig. \ref{fig1}a, while a magnified view of the junction's edge is shown in Fig. \ref{fig1}b. Three devices of various MTJ sizes to maintain a $\sim$5x aspect ratio (length to width) are referred to as L350, L400, and L600, with MTJ lengths of 350 nm, 400 nm, and 600 nm, and MTJ widths of 70 nm, 90 nm, and 120 nm, respectively, depicted in Fig. \ref{fig1}c. A 60 nm diameter MTJ is used as the control, depicted in Fig. \ref{fig1}d. The device configurations exhibit an RA product of 10.9 $\Omega$-$\mu$m$^2$, extracted from the linear fit of MTJ's R$_{min}$ versus 1/A shown in Fig. S1 The average tunnel magnetoresistance (TMR) values are 80.2\%, 44.8\%, 39.1\%, and 30.5\% for the control, L350, L400, and L600, respectively. This shift in TMR from control to MTJ artificial synapses can be explained by the fabrication process being optimized for conventional MRAM, shown in Fig. \ref{fig1}b, but still providing a sufficient readout margin.

SOT-induced magnetization switching was performed with IP magnetic fields ranging from -300 Oe to 300 Oe in 50 Oe increments, with 25 cycles executed at each field. The IP field, applied parallel to the current, served to break the symmetry of the SOT-induced torque, thereby enabling deterministic switching. A constant 200 Oe magnetic field is also applied in the $\hat{z}$ direction to compensate for the stray field from the synthetic anti-ferromagnet (SAF). All measurements were performed using a pulse measure scheme with the parameters shown in Table \ref{table1}. Pulse amplitude encoding is incorporated, with the voltage amplitude increasing with each subsequent pulse in a given ramp direction. The MTJ's resistance is measured by applying a sense current through the MTJ during the inter-pulse interval, 1 ms, to ensure a stable readout of the MTJ's state. During the applied SOT pulse, a compensation voltage (V$_{Comp}$) of 50\% of the amplitude of the SOT pulse was applied to the top of the MTJ while one of the bottom electrodes is grounded to ensure that most of the current flows through the SOT line, minimizing the voltage drop across the stack.   

\begin{table}[htbp]
    \centering
    \caption{SOT-induced magnetization switching pulse measure scheme parameters.}
    \label{table1}%
    \begin{tabular}{@{}cccc@{}}
    \toprule
     & Sense Current ($\mu$A) & Max Amplitude (V) & Pulse Width (ns) \\
    \midrule
    Control    & 3 & 0.6 & 10 \\
    L350       & 10 & 1.5 & 10 \\
    L400       & 10 & 1.5 & 10 \\
    L600       & 30 & 1.7 & 50 \\
    \botrule
    \end{tabular}
\end{table}


\section{Results}\label{results}

\begin{figure}[htbp]
    \centering
    \includegraphics[width=1\textwidth]{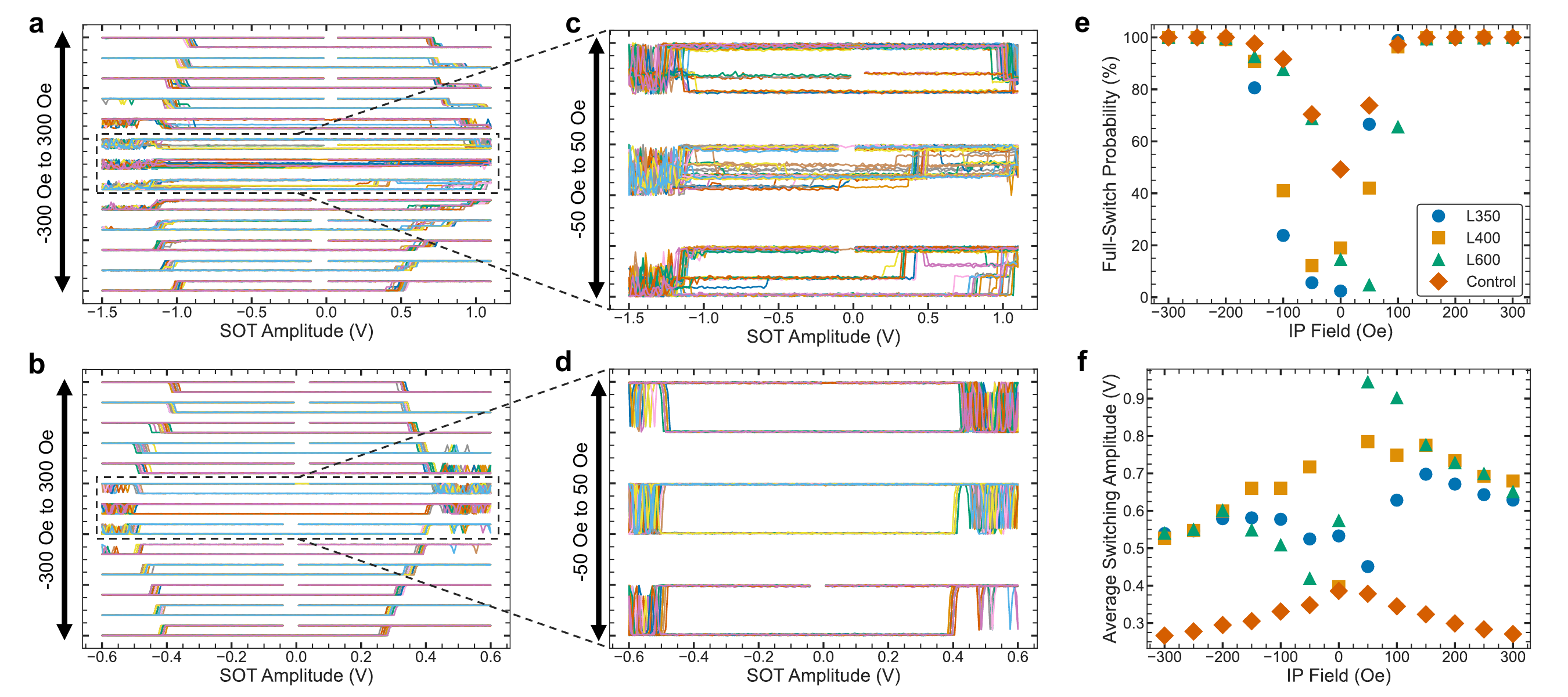}
    \caption{\textbf{SOT-induced magnetization switching loops and switching  analysis of the MTJ artificial synapses.} (\textbf{a}) Normalized SOT switching loops for the MTJ artificial synapse, L400, at varied IP fields. (\textbf{b}) Normalized SOT switching loops for the control at varied IP fields. (\textbf{c}) Normalized SOT switching loops for L400 when the IP fields are -50 Oe, 0 Oe, and 50 Oe. (\textbf{d}) Normalized SOT switching loops for the Control when the IP fields are -50 Oe, 0 Oe, and 50 Oe. (\textbf{e}) Probability for each MTJ artificial synapse to achieve a complete switch for all 25 cycles. Blue/circles, yellow/squares, green/triangles, and red/diamonds represent the devices L350, L400, L600, and the control, respectively. (\textbf{f}) Average switching amplitude achieved at each IP field across all subset of devices.}
    \label{fig2}
\end{figure}

To evaluate the switching characteristics of the MTJ artificial synapses, SOT-induced magnetization switching loops were measured on a subset of 20 devices for each configuration. Representative switching loops for the L400 MTJ artificial synapse and the control device are shown in Fig. \ref{fig2}a and Fig. \ref{fig2}b, respectively, while additional loops for all configurations are provided in Section 2 of the Supplementary. From these plots, two key trends can be observed: reducing the IP field reduces deterministic switching, and the MTJ artificial synapses develop stable multi-weight switching, while the control remains binary. This behavior is further emphasized through Figs. \ref{fig2}c,d which compares switching loops measured at IP fields of $-50$, $0$, and $50$ Oe for L400 and the control, respectively. The control device exhibits abrupt switching followed by expected back-hopping at higher SOT amplitudes \cite{ren2021strongly}. In contrast, L400 no longer switches directly between parallel and antiparallel states. Instead, multiple stable resistance levels are accessed during the switching loops, producing MW behavior. Rather than switching abruptly, the resistance evolves through several stable intermediate configurations, indicating the formation of DWs within the MTJ due to the non-uniform voltage distribution V$_{Comp}$.

The trends observed in the  switching loops are reflected in the analysis of the measured devices quantifying the probability of achieving a full switch and the average switching amplitude of the SOT switching loops, shown in Fig. \ref{fig2}e,f. A full switch is defined by comparing the TMR obtained during each cycle with the TMR extracted from the corresponding device's field loop, shown in Fig. S5 in the Supplementary. Fig. \ref{fig2}e displays the probability of achieving a full switch switch that decreases as the IP field is reduced for all devices. This reduction is particularly pronounced for the MTJ artificial synapses, where the probability falls below 20\% near zero IP field, whereas the control falls to 50\%. These results indicate that the IP field provides the symmetry breaking required for deterministic SOT switching. The average switching amplitude, shown in Fig. \ref{fig2}f, shows the average SOT voltage required to produce any stable resistance change, regardless of whether a complete switch is achieved, thus capturing transitions into intermediate resistance states. As the IP field approaches zero, the MTJ artificial synapse's average switching amplitudes reduce, indicating that a stable resistance change occurs at lower voltages despite complete switching becoming increasingly unlikely. In contrast, the control device's switching amplitude increases at lower IP fields, consistent with its predominantly binary switching behavior. Together, the trends from the full switch probability and average switching amplitude suggest that at little to no IP fields, the free-layer magnetization remains largely perpendicular during the SOT pulse, promoting DW formation. These DWs inhibit complete magnetization reversal while stabilizing multiple intermediate resistance states, enabling the observed MW switching behavior.

\section{Simulations}\label{simulations}

\begin{figure}[htbp]
    \centering
    \includegraphics[width=1\textwidth]{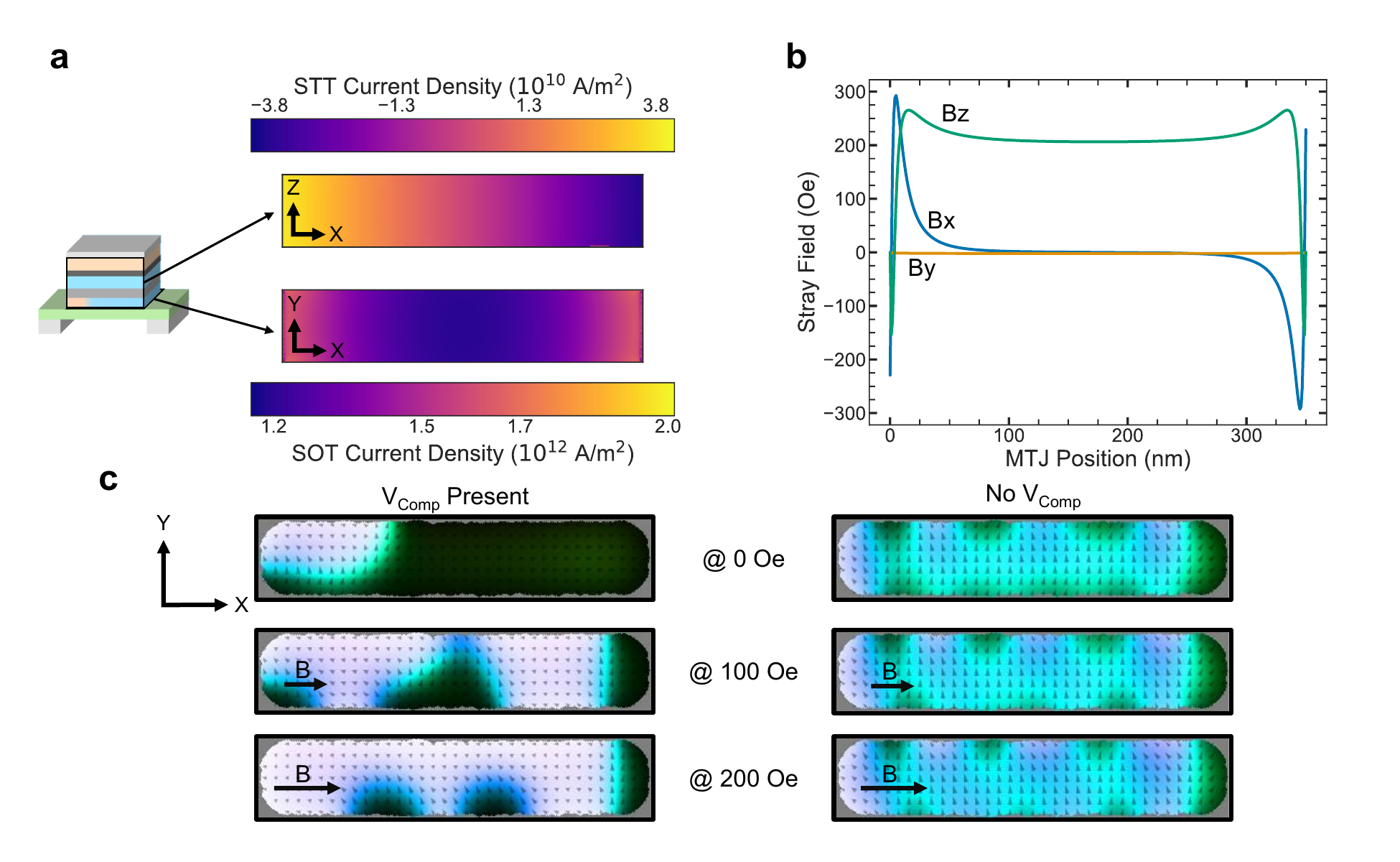}
    \caption{\textbf{Simulations investigating DW nucleation in the MTJ artificial synapses.} (\textbf{a}) COMSOL simulation of the SOT surface current density in the W heavy metal and the STT current density through the MTJ under SOT pulsing. (\textbf{b}) Stray field distribution originating from the hard and reference layers of the MTJ. (\textbf{c}) Mumax3 simulation showing the top down magnetization of the MTJ during the SOT pulses at varying IP fields with a V$_{Comp}$ present (left) and no V$_{Comp}$ resulting in a uniform SOT current density (right), both with stray field from the RL and SAF. Blue and green signifies IP magnetization and black and white signifies OOP magnetization.}
    \label{fig3}
\end{figure}

To assess the origin of DW behavior in MTJ artificial synapses, we performed simulations using COMSOL and Mumax3 \cite{COMSOL, Mumax3}. Several mechanisms could contribute to the formation of domains, including variations in voltage potential, stray field, edge roughness, and grain boundaries. As described in Section \ref{device_structure}, V$_{Comp}$ is applied to the top of the MTJ during the SOT pulse at 50\% of the pulse's amplitude while one of the bottom electrodes is grounded. For smaller MTJs, the voltage potential across the junction can be assumed to be uniform; however, in MTJ artificial synapses the potential becomes non-uniform due to the $\sim$5x aspect ratio. Using COMSOL, the resulting current distribution in the free layer (J\textsubscript{STT}) and on the surface of the heavy metal (J\textsubscript{SOT}) was simulated in the L350 device, with the results shown in Fig. \ref{fig3}a. When the SOT current pulse is present, an uneven current density distribution is produced, introducing a small STT leakage current that flows through the edges of the MTJ. The corresponding current density magnitudes are on the order of 1E10 A/m$^2$ for STT at the corners of the MTJ and 1E12 A/m$^2$ for SOT. Thus, the leakage current from the non-uniform voltage potential likely contributes to the formation of DWs in the MTJ artificial synapses.

Stray fields also influence magnetodynamics, particularly at smaller scales. To capture these effects, we simulate the stray fields from the magnetic stack using Mumax3. In these simulations, contributions from both the reference layer and the SAF are included. The resulting stray field profile in the CoFeB free layer is shown in Fig. \ref{fig3}b. The maximum stray field is approximately 300 Oe in the $\hat{x}$ direction. However, these fields, while capable of affecting a DW due to it's chirality, are localized at the edges of the MTJ. If the nucleation of the DWs were influenced by the stray field, it would occur at $\sim$10 nm from the edge of the MTJ, leading to a resistance change of only 1.6\% in the L350 device. Such a small variation to the MTJ's resistance is not significant compared to the measured data, indicating that the stray field is unlikely to be the primary contributing factor to the formation of a DW.

To test the impact of the leakage current and stray field, micromagnetic simulations of the device were performed. These simulations incorporate the corresponding outputs for STT, SOT and stray fields from Fig. \ref{fig3}a-b where exact methods and parameters are discussed in Section 5 of the Supplementary. An external field in the $\hat{z}$ direction was added to the simulation to follow similar methods used for the measurements of the experimental MTJ artificial synapses. Furthermore, these simulations were performed at 0 K to gain an understanding of how domains form without any thermal effects. The resulting output tracks the magnetization of the MTJ's free layer during the SOT pulse under varying IP fields when V$_{Comp}$ is applied to the top electrode introducing an STT leakage current and when V$_{Comp}$ is not included, resulting in a uniform SOT current, shown in Fig. \ref{fig3}c. These results indicate that with no field and V$_{Comp}$ present, the STT and SOT currents contribute constructively to domain formation on a single side of the MTJ. However, as the field increases, the switching becomes more deterministic because the IP fields introduce a symmetry breaking component to the SOT, overpowering the non-uniformity from the leakage current. When V$_{Comp}$ is not included, the SOT current along the heavy metal is uniform and the domains do not form. Instead, the magnetization of the free layer aligns in the in-plane direction and is not affected by the IP field; further supporting the idea that the MTJ's stray field alone is not able to form domains. These simulations support the experimental trends that demonstrate that with the non-uniform voltage potential due to V$_{Comp}$, the free-layer's magnetization remains largely perpendicular during an SOT pulse at zero field, promoting the nucleation of domains.


\section{Analysis}\label{analysis}

\begin{figure}[ht]
    \centering
    \includegraphics[width=1\textwidth]{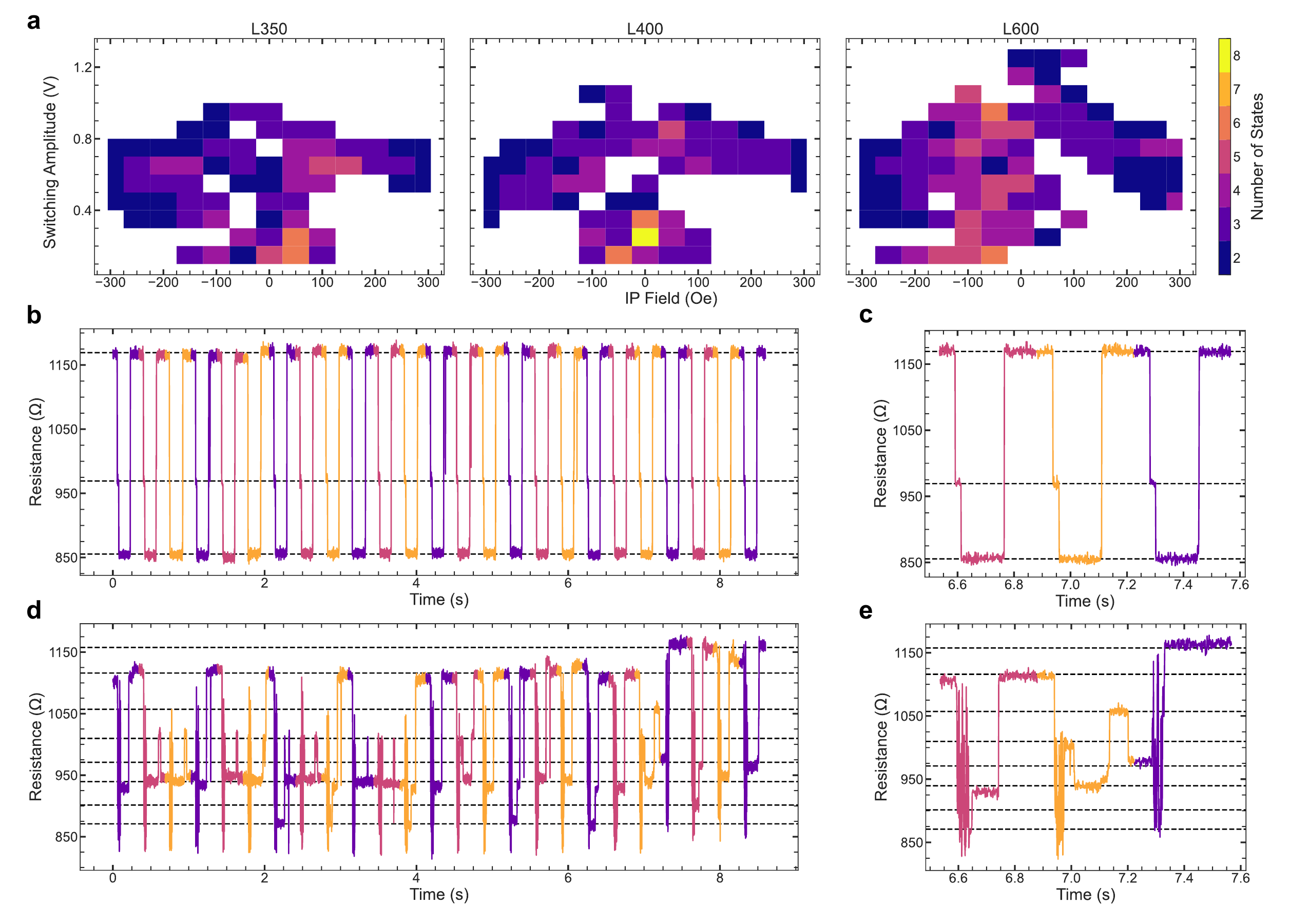}
    \caption{\textbf{MW and Temporal behavior of the MTJ artificial synapses showings the different conductance states and their programming capabilities.} (\textbf{a}) Switching amplitude map for the respective MTJ artificial synapse as a function of the IP Field. Color indicates the maximum number of distinct resistance states observed at each coordinate, ranging from 2 (purple) to 8 (yellow). (\textbf{b}) Temporal behavior of an example L400 device at 150 Oe showing three resistance states. (\textbf{c}) Temporal behavior of the L400 device at 0 Oe showing eight resistance states throughout the 25 cycles. (\textbf{d}) Three cycles of the temporal behavior of the L400 device at 150 Oe. (\textbf{e}) Three cycles of the temporal behavior of the L400 device at 0 Oe.}
    \label{fig4}
\end{figure}

To analyze how switching amplitudes translate into MW behavior, Fig. \ref{fig4}a evaluates the number of achievable resistance states for each device as a function of IP field and switching amplitude. For each device, the number of states is defined by the total set of distinct, achievable resistance states, including both the high and low resistance states of the MTJ. A resistance is considered addressable if it is reproducibly reached after five or more consecutive SOT pulses within a single cycle, demonstrating that it can be reliably accessed during synaptic programming.

At high IP fields, the switching amplitudes cluster within a tight range and the devices exhibit only a small number of states, indicating predominantly binary behavior. However, as the IP field is reduced, the distribution of the switching amplitude widens, and some devices exhibit switching amplitudes below the junction's thermal stability threshold. Therefore, only switching amplitudes $>$ 0.1 V are considered to represent thermally stable state transitions. These low switching amplitudes indicate that a DW remains in the device within programming cycles, allowing subsequent pulses to translate an existing DW rather than nucleating a new one. These experimental observations are consistent with the simulation results, which predict the formation of DWs as the IP field is reduced. Consequently, each device configuration produces at least six resistance states for two devices at IP fields of 50 Oe, 0 Oe, and -50 Oe for L350, L400, and L600, respectively. Notably, L400 reaches eight resistance states in one device at 0 Oe, corresponding to a three-bit artificial synapse.

To examine how these states can be harnessed for programming, Figs. \ref{fig4}b-e show representative temporal switching behavior for the L400 MTJ artificial synapse, which achieved eight resistance states at zero IP field. The time-axis for these plots are reconstructed by summing the pulse widths for the respective devices where the pulse widths are shown in Table \ref{table1}. Each programming cycle is shown in a different color, with additional temporal behavior for the other configurations provided in Section 4 of the Supplementary Information. At an IP field of 150 Oe, Fig. \ref{fig4}b, the switching is largely deterministic: each cycle produces a full switch, occasionally accompanied by an intermediate resistance state. A closer examination of three cycles, Fig. \ref{fig4}c, shows that this intermediate state appears consistently within each selected cycle, although it is not present in every programming cycle and occurs only during one polarity of the SOT switching loop. The repeated appearance of these intermediate states indicates that the DW is retained within the programming cycle rather than being annihilated, suggesting a high probability that the DW becomes pinned at the same location. In contrast, at zero IP field, Fig. \ref{fig4}d, the switching behavior changes dramatically. Rather than consistently achieving a full switch, the DW propagates to multiple locations across the MTJ artificial synapse, producing eight distinct resistance states that occur with varying probability over repeated programming cycles. Three representative cycles are shown in Fig. \ref{fig4}e, showing that while programming events become increasingly stochastic, the number of accessible resistance states increases. Consequently, although the selection of the resistance state requires repeated programming attempts, the increase in number of states provides a larger set of programmable levels for neuromorphic applications. 

A comparison of these devices with other wafer-scale technologies capable of analog or multilevel synaptic responses is provided in Table \ref{table2}, highlighting the combination of multistate operation, programming characteristics, and device-level performance achieved in this work. Write noise is defined as the standard deviation of the resistance readings within each unique resistance state, normalized to that state's mean value and expressed as a percentage, then averaged across all states. Device-to-device (D2D) variation is defined as the coefficient of variation (standard deviation normalized to the mean, expressed as a percentage) of the low- and high-resistance states, computed independently across all 20 devices. It can be seen that the write noise for the MTJ artificial synapses is low due to the stability of the magnetization once it settles into a stable configuration. Furthermore, these devices achieve fast write speeds and low write energy relative to the number of states they support.

\begin{table}[htbp]
    \centering
    \caption{Comparison of experimentally measured characteristics of the control and multi-weight SOT-MTJ devices.}
    \label{table2}
    \begin{tabular}{@{}lccccccc@{}}
    \toprule
    Device & Write Speed & Write Energy & Write Noise & States & D2D Variation \\
    \midrule
    L350    & 10 ns & 1.5 pJ & 0.44\% & $\geq 6$ & 5.1\% \\
    L400    & 10 ns & 1.5 pJ & 0.62\% & $\geq 8$ & 5.7\% \\
    L600    & 30 ns & 1.7 pJ & 0.77\% & $\geq 6$ & 5.6\% \\
    Control & 3 ns  & 0.6 pJ & 0.42\% & 2 & 11.1\% \\
    SOT MRAM \cite{Liu_2025} & 2 ns & 2 pJ & 0.107\% & 2 & $\sim$10\% \\
    STT MRAM \cite{Jung_2022} & 90 ns & 27 pJ & -- & 2 & -- \\
    RRAM \cite{bennett2019wafer} & 10 ns & 100 pJ & $\sim$11.5\% & 8 & $\sim$23.1\% \\
    RRAM \cite{park2024multi} & 5 ms & 787.5 pJ & $\sim$1\% & 16 & -- \\
    RRAM \cite{choi2025wafer} & 80 $\mu$s & 400 pJ & $\sim$1\% & 16 & $\sim$29\% \\
    ECRAM \cite{Faruque_2025} & $\sim$10 ms & -- & -- & 128 & $\sim$0.08\% \\
    ECRAM \cite{chen2023open} & 100 ns & 3.5  $\mu$J & $<$1\% & $\sim$90 & $\sim$1.1\% \\
    PCRAM \cite{song202112} & 50 ns & 1 pJ & $\sim$5\% & 12 & $\sim$9\% \\
    PCRAM \cite{le202364} & $<$100 ns & $\sim$2 nJ & $\sim$8\% & $\sim$12 & -- \\
    \bottomrule
    \end{tabular}
\end{table}


\section{Applications}\label{applications}

\begin{figure}[htbp]
    \centering
    \includegraphics[width=1\textwidth]{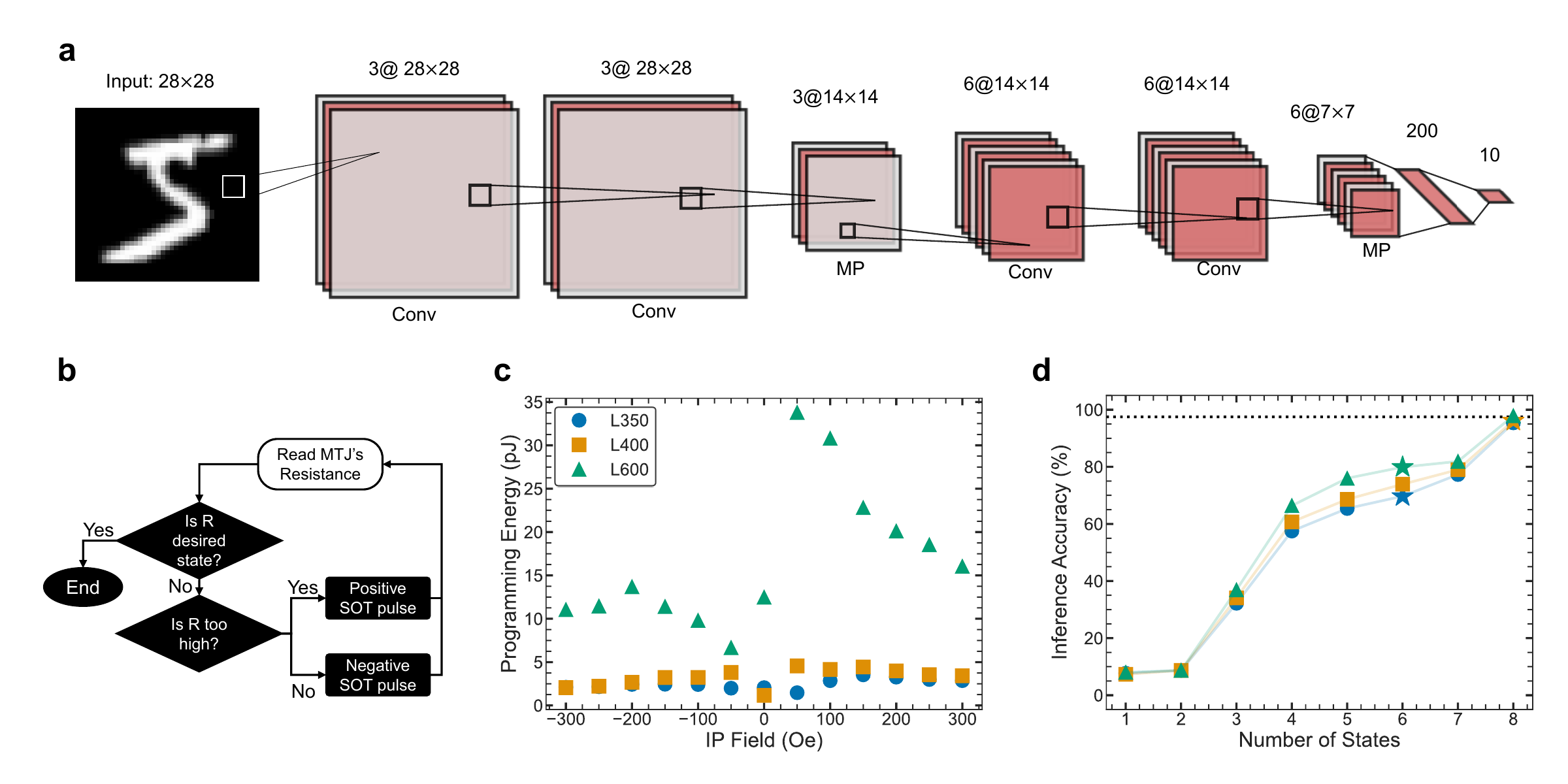}
    \caption{\textbf{Programming and inference utilizing wafer-scale MTJ artificial synapses.} (\textbf{a}) Neural network architecture used for MNIST inference. (\textbf{b}) Iterative programming algorithm tailored for MTJ artificial synapses. (\textbf{c}) Programming energies of the MTJ artificial synapses for a single cycle. Blue/circles, yellow/squares, and green/triangles represent the devices L350, L400, and L600, respectively. (\textbf{d}) Inference Accuracy based on the number of resistance states in the MTJ artificial synapses implemented in a crossbar array. The  starred points represent a crossbar full of MTJ artificial synapses with the maximum number of measured resistance states.}
    \label{fig5}
\end{figure}

To evaluate the suitability of these devices as synaptic elements, we modeled a crossbar array using the experimentally measured device data to test performance on a pretrained MNIST dataset, with the architecture shown in Fig. \ref{fig5}a. The maximum number of resistance states experimentally achieved is eight for the L400 configuration, corresponding to three weight bits. CrossSim, a crossbar array simulator, is used to map resistance state distributions to synapse weights in a differential bit-cell architecture \cite{CrossSim}. This configuration supports up to four bit signed weights during inference using the MTJ artificial synapses because each bit cell uses two devices to represent positive and negative weights.

The MTJ states can be addressed by using a standard iterative programming algorithm similar to that used for RRAM, shown in Fig. \ref{fig5}b \cite{Esmanhotto_2020, Esmanhotto_2022}. This algorithm repeatedly reads and writes the device until the target resistance state is reached when using pretrained models that quantize the weights based on the number of resistance states of the synapse. A maximum of 25 programming cycles is assumed, as each accessible resistance state was experimentally observed at least once within 25 programming cycles. The programming energy for a single cycle is shown in Fig. \ref{fig5}c. The programming energy for the different MTJ artificial synapses that achieved the maximum number of resistance states are 1.47 pJ, 1.17 pJ, and 6.67 pJ for L350, L400, and L600, respectively. Using the iterative programming algorithm assuming a full 25 cycles is needed, the corresponding worse-case programming energies increase to 36.8 pJ, 29.3 pJ, and 166.8 pJ. 

Inference was performed using quantized weights according to the number of resistance states measured. Three accuracy metrics were examined: the effect on the number of resistance states, the accuracy when all synapses are assumed to reach their maximum number of states, and the impact of device-to-device variations. These results are summarized in Fig. \ref{fig5}d. The dashed black line shows the baseline accuracy of 97.5\% without quantization. As the number of states increases, the accuracy improves because the synapses become more expressive, with eight-state devices approaching the baseline. The starred points show the accuracy assuming all devices in the specific configuration can achieve the maximum number of measured resistance states, yielding 69.7\%, 96.1\%, and 79.9\% for L350, L400, and L600, respectively. When device-to-device variations are included, the maximum accuracy yielded 48.6\% with the L350 configuration at an IP field of 50 Oe. Evaluations for each configuration with the variations in IP field are shown in Fig. S9 in the Supplementary. Overall, these results indicate that designing devices to achieve a large and consistent number of resistance states significantly improves inference performance and enables deployment of complex datasets on the MTJ array compared to binary-only MTJs.


\section{Conclusion}\label{conclusion}

We have demonstrated that elongated MTJs can serve as artificial synapses. Statistical analysis across 128 devices on a 300 mm wafer revealed that the number of resistance states increases as the symmetry-breaking IP field decreases. Each configuration achieved at least six resistance states, with two devices reaching six states and one device reaching eight states in the 400 nm long device. The observed trends highlight differences in programming behavior where the programming is increasingly stochastic but with a larger number of resistance states. Simulations support the hypothesis that the intermediate states are a result of DW nucleation and propagation because of the dual current in both the SOT track and MTJ thin film stack. The measured statistics were evaluated on a modeled neural network crossbar array, showing that the measured eight state synapse can achieve inference accuracies of 96.1\%, and device-device variations achieving inference accuracies up to 48.6\%. Future investigations to lower the device-to-device variation at lower IP fields would lead to notable gains in inference accuracy. These results show the viability of higher aspect ratio MTJs as scalable synaptic elements for neuromorphic systems. With further improvements in read window, the high endurance combined with the presented multi-weight behavior would make MRAM a strong contender for nonvolatile memory based crossbar arrays.


\section{Methods}\label{methods}

\subsection{Fabrication}

The devices were fabricated using IMEC's 300 mm MRAM pilot line. The SOT-MTJ stack was deposited on preprocessed wafers featuring W-VIA bottom electrodes which are embedded in SiO$_2$. The stack was grown in situ at room temperature by physical vapor deposition using a 300 mm EC7800 Canon-Anelva cluster tool, followed by a 30-min anneal at 300 °C under a 1 T perpendicular magnetic field. The top-pinned W/CoFeB MTJ stack consisted of Substrate/ W(3.5)/ Co 20 Fe 60 B 20 (0.9)/MgO ($\sim$1)/Co 17.5 Fe 52.5 B 30 (1.1)/W(0.3)/ Co(1.2)/Ru(0.85)/Co(0.6)/Pt(0.8)[Co(0.3)/Pt(0.8)] 6 Ru(5), with all thicknesses in nanometers. MTJ pillars were defined using 193 nm immersion lithography and ion-beam etching at normal incidence with an optimized etch-stop condition. Copper top and bottom electrodes were then formed using a dual-damascene process to complete the electrical contacts.

\subsection{Measurement}

Measurements were conducted using the Hprobe RF Spin system at room temperature and atmosphere. All measurements were performed using a pulse measure scheme where resistance was measured using a Keithly DMM 7510 and the current was sourced through a Keithley SMU 2450.


\section{Acknowledgements}\label{acknowledgements}

N.Z. and J.A.I acknowledges UT Austin Discovery to Impact and Proof of Concept awards. This work has been enabled in part by the NanoIC pilot line. The acquisition and operation are jointly funded by the Chips Joint Undertaking, through the European Union’s Digital Europe (101183266) and Horizon Europe programs (101183277), as well as by the participating states Belgium (Flanders), France, Germany, Finland, Ireland and Romania. For more information, visit nanoic-project.eu. This work is also supported by imec’s industrial affiliation program on MRAM devices. D.G. acknowledges FWO-Vlaanderen for a Strategic Basic Research PhD fellowship No. 1SHEV24N.

\section{Author Contributions}\label{author-contributions}

N.Z. measured the devices, analyzed the data, and wrote the manuscript with support from D.G. and V.K. V.D.N. guided the experimental setup and measurements. J.C., A.P., and R.C. provided feedback on the manuscript and contributed to the discussion of the results. J.A.C.I. and S.D. led the team in the design, measurement, analysis, and writing. The manuscript was written with input from all authors and all authors discussed the results.

\section{Data Availability}\label{data-availability}

The data presented in this study are available from the corresponding author on reasonable request with an estimated 72 hr response time.

\backmatter

\bibliography{sn-bibliography}
\end{document}